# Ultra-Thin Silver Films obtained by Sequential Quench-Anneal Processing

S.B. Arnason[a], A. F. Hebard[b]

(a) Department of Physics, University of Massachusetts Boston, 100 Morrissey Blvd Boston, MA, 021225, USA, stephen.arnason@umb.edu, corresponding author

(b) Department of Physics, University Of Florida, Gainesville, FL, 32611, USA, afh@phys.ufl.edu

## Abstract

We have used the "two-step" growth technique, quench condensing followed by an anneal, to grow ultra thin films of silver on glass substrates. As has been seen with semiconductor substrates this process produces a metastable homogeneous covering of silver. By measuring the *in situ* resistance of the film during growth we are able to see that the low temperature growth onto substrates held at 100 Kelvin produces a precursor phase that is insulating until the film has been annealed. The transformation of the precursor phase into the final, metallic silver film occurs at a characteristic temperature near 150K where the sample reconstructs. This reconstruction is accompanied by a decrease in resistance of up to 10 orders of magnitude.

## Introduction;

The potential usefulness of ultra-thin homogeneous metallic films in technology and basic science makes any new deposition technique that can provide such films is of great interest. One of the more exciting recent developments in this field has been the realization that the "two-step" growth process, quench condensing of a metallic film onto a cryogenically cooled substrate, followed by annealing to room temperature, can produce epitaxial thin films of silver on semiconductor substrates.(Smith, Chao, Niu and Shih 1996; Chao, Zhang, Ebert and Shih 1999) This result is somewhat counterintuitive since the growth process is far from equilibrium, and one normally expects epitaxy for quasi-equilibrium growth as is the case in Molecular Beam Epitaxy.

The films that result from the application of this two-step process to growth on single crystals of III/V semiconductors are not only epitaxially related to the substrate, but are atomically flat, growing at specific, "magic" thicknesses. Excess material forms islands atop the atomically flat layer. Insufficient material produces voids, plunging through several atomic layers to the substrate or to a monatomic wetting layer. The thickness of these films is so uniform that the resulting quantization of the electronic states, due to confinement along the direction normal to the substrate surface, is observable in photoemission spectroscopy,(Neuhold and Horn 1997; Matsuda, Yeom, Tanikawa, Tono, Nagao, Hasegawa and Ohta 2001) where it modulates the density of states near the Fermi level as a function of energy.

It is thought that the stability of the films at magic thicknesses is a consequence of this quantum size effect.(Zhang, Niu and Shih 1998; Luh, Miller, Paggel, Chou and Chiang 2001) A commensuration of the wavelength of the component of the electronic wave function perpendicular to the substrate with the thickness of the film creates a lower total energy state at the "magic" thicknesses. This "electronic growth" model gives good agreement with experimental observations. Another parameter critical to this picture is the boundary condition of the electronic states at the metal semiconductor interface. A leakage of electron density into the semiconductor lowers the total energy of the system.

The stability of the microstructure of a thin film is usually thought of as a consequence of phenomenological energy scales associated with the wetting interaction, cohesive energy, and elastic strain. A material that wets another will form a two dimensional film on that surface, Frank-van der Merwe growth. One that does not will break up into three-dimensional quasi-hemispherical grains, Volmer-Weber growth, whose geometry is dominated by the cohesive energy. Finally, strain, resulting from lattice mismatch can inhibit two dimensional growth.

In the electronic growth model there is another energy scale in addition to those traditionally considered. The quantum size effect can lower the total energy of the film if it produces a net shift of spectral weight in the density of states to deeper binding energy. This energy is associated with the formation of a quantum well as the consequence of the thin coverage. If the wavelength of the resulting electronic states, taking into account the phase shifts at the interfaces, is commensurate with the thickness of the film there is a net lowering of the total energy associated with the electronic states in the film.

The "two-step" growth process also produces remarkable films on polycrystalline and amorphous substrates.(Arnason and Hebard 1997) While, in this case, there is obviously no epitaxy, we have used "two-step" growth on glass substrates to produce ultra-thin, homogeneous films of silver.(Arnason and Hebard 1997) AFM shows these films to be extremely flat, but not having a single dominant thickness. This suggests that while the quantum size effect may be important here it is not the only factor of significance in determining the final microstructure. Importantly, this also suggests that much of the uniqueness in two-step growth may be a consequence of the quenched kinetics rather than the quantum energies involved.

This idea that there is a kinetic constraint on the microstructure is also supported by the metastability of the two-step films. Our films on glass are stable at room temperature but reconstruct into isolated three dimensional droplets at about 50°C, the agglomeration temperature, $T_a$. Shih's group sees a similar reconstruction for their films on GaAs, but at 670°K.(Yu, Jiang, Ebert, Wang, White, Niu, Zhang and Shih 2002)It is likely that the electronic growth mechanism should be thought of as an energy barrier between the homogeneous ultra-thin films and the thermodynamically stable structures that growth techniques closer to equilibrium produce. Clearly the energy associated with the quantum well state is important, but these films are fundamentally metastable. It is only because we have traveled along a kinetically constrained path of growth that we are able to find this local energy minimum.

Given the importance of the growth kinetics it is necessary that we understand the physics of the low temperature phase if we are to understand two-step growth in detail. To date the work in this field has primarily focused on the final product, the unique film that results from "two-step" growth. This paper presents results on the precursor phase and, equally importantly, about the transformation that it undergoes during the annealing process. To accomplish this we measure the *in situ* resistivity throughout the entire growth. This gives us a measurement of a property of the material which can be tracked through the actual transition. We then compare the resulting material to films grown without constrained kinetics by measuring the resistivity as a function of temperature, *ex situ*.

**Experimental Details**

Our films are grown in high vacuum, base pressure 2 X 10$^{-8}$ Torr. The silver, 99.999% purity, is evaporated from an alumina crucible at a rate of approximately 1 Å per minute. The substrate is cooled by thermally anchoring it to a liquid nitrogen reservoir. Its temperature is measured by a type J thermocouple mounted on the surface of a second substrate in an attempt to mimic the imperfect thermal contact. The temperature during deposition is approximately 100K, though variation of 1-2° K due to radiative heating by the Silver source is observable. To minimize cryopumping of contamination onto the substrate surface before deposition the substrate was kept at room temperature till the source was stabilized at the desired temperature and then quickly cooled to growth temperature, ~5 minutes.

The thin films are deposited through a shadow mask of a standard "Hall Bar" onto a substrate with pre-deposited gold\chromium electrodes. The resistance is measured by applying a voltage and measuring the resulting current with an electrometer, with remote voltage sensing on a second set of contacts so as to minimize the effects of contact resistance. The sign of the applied voltage is alternated and averaged to compensate for any DC offsets. The frequency of voltage reversal is chosen so that we can ignore currents due to capacitative charging which are significant when the samples are highly resistive. As the sample resistance decreases the excitation voltage is varied through a range from 2 volts to 2 millivolts during the measurement to keep a good signal to noise ratio.

After growth the resistivity as a function of temperature of the films is measured *ex situ* in a liquid Helium cryostat. Here the measurements are made with an AC resistance bridge at 17 Hz using an excitation voltage of 2 mV. The measurement temperatures ranged from 2 to 300 K.

**Results**

Figure 1 shows the resistance as a function of coverage (the mass of deposited material per unit area, normalized by density) for samples as they grow at 100° K and at room temperature. At 100° K there is a steady monotonic decrease of resistance with increasing thickness, suggesting that the morphology of the film is consistent throughout the growth. At a thickness of about 160 Ångstroms we doubled the growth rate and there is a noticeable change is the slope of resistance as a function of thickness but the dependence is still monotonic. This is to be contrasted with room temperature growth where the resistance initially varies exponentially with thickness, changing to a power law at the percolation threshold, and finally crossing over to 1/(thickness) dependence when the film becomes homogeneous.(Arnason and Hebard 1998) The room temperature growth occurs initially as nucleation of isolated clusters, which grow by the addition of atoms from the arriving flux that diffuse over the surface. The sudden drop in resistance occurs when the islands reach a density equivalent to the percolation threshold. Finally, at thick enough coverage, the film becomes effectively homogeneous within the plane of the substrate and the resistivity scales with the inverse of the thickness, as one would expect from simple geometric considerations within a Fuchs-Sondheimer picture. (Fuchs 1938)

If we terminate the growth of a quench condensed film at a coverage of 20 to 60 Å and measure the resistance during the anneal to room temperature we see a dramatic change of resistance at a characteristic temperature of about 150 K, henceforth referred to as $T_{tr}$ (see figure 2). Below $T_{tr}$ the sample is an insulator with activated conduction. In this region the resistance is reversible as a function of temperature. When the sample warms to $T_{tr}$, an irreversible reconstruction of the film begins. As a result of this reconstruction, the sample transforms from an ultra-thin insulator into an ultra-thin homogeneous metallic film. Quenching of the reconstruction can be achieved by rapidly cooling the sample before it has completely reconstructed to its final state (see figure 3). If we perform this quench before the sample has annealed to a sheet resistance lower than ~20 kΩ, then the low temperature resistivity as a function of temperature is activated. If we arrest the anneal after the sheet resistance has fallen below this value, the temperature dependence is metallic.

Allowing the sample to anneal to room temperature results in a film which is metastable, showing a reversible dependence of resistivity on temperature as long as the temperature does not go above $T_a$, where the sample will irreversibly reconstruct into isolated hemispherical islands of silver. For our samples on glass this temperature is typically ~350° K.

The morphology of the films that are raised above this second characteristic temperature, as measured by atomic force microscopy, is equivalent to that of films grown at room temperature whose growth was terminated at the same coverage, showing isolated grains of silver. The two-step films that are kept at room temperature show a homogeneous coverage with no grain structure visible at length scales within the resolution of the instrument.(Arnason and Hebard 1997) These films are also relatively stable in an inert atmosphere, showing no change in the resistivity as a function of time over several weeks.

Further insight into the microstructure can be gleaned from the *ex situ* measurement of resistivity as a function of temperature. Figure 4 shows resistivity as a function of temperature for three films of silver; a two-step film, a film grown at room temperature where the growth has been terminated in the percolative, scaling regime, and a room temperature grown film that has been grown to the point of showing resistivity scaling with inverse thickness. While all three films show relatively high residual resistivity, indicating that they are disordered, the percolating silver film also shows an enhanced temperature dependence of resistivity. This enhanced temperature dependence is a consequence of the conversion of the resistance as a function of temperature data to resistivity by scaling it with the effective thickness of the film, but not taking into account the inhomogeneous morphology of the percolating cluster. The resistivity for the percolating film can be mapped onto that of the homogeneous film by dividing the data by a single geometric factor, a factor that is a measure of the fractal geometry of the cluster.(Arnason and Hebard 1998) Thus, the enhancement of the temperature dependence of resistivity away from that of bulk silver can be used to measure the geometric inhomogeneity of the system. The data for the two-step film has enhanced residual resistivity but temperature dependence indistinguishable from the homogeneous film. This observation indicates that the resisitivity can be used as a measure of sample homogeneity and that the two-step films are extremely homogeneous.

**Discussion;**

Clearly the most compelling data presented in this paper is that of the resistive signature of the morphological reconstruction shown in figure 2. The data shown here is typical. All two-step films in the thickness range studied show this transition. There is always a characteristic temperature at which it begins, but as the samples become thicker the breadth of the transition in temperature increases. The existence of this sharp transition suggests that there is a very specific energy barrier that must be overcome for the films to reconstruct. Besides the sharpness of the transition, the magnitude of the change is also quite striking. A change of resistance of ten orders of magnitude indicates that there is a significant change in the nature of the silver film. The precursor phase, which anneals into the ultra thin metallic silver film, is a good insulator. The final state is a good metal.

The existence of this specific relatively low energy scale for transformation is quite surprising. All of the characteristic energy scales associated with the mobility of silver are significantly higher. For example the activation energy for grain growth, which is similar in magnitude to the energy scale associated with surface diffusion, has been found to be on the order of 530 meV.(Dannenberg, Stach, Groza and Dresser 2000) The lowest energy scales that we are aware of have been identified in Molecular Dynamics simulations of diffusion processes on the (111) facets of small silver clusters with activation energies as low as 70 meV.(Baletto, Mottet and Ferrando 2000) Even this is several times larger than the 13 meV that is equivalent to 150°K.

A possible model of this transition is that islands of metallic silver are nucleating inside of a host of the insulating precursor. The crossover sheet resistance corresponds to a percolation of these islands into a contiguous backbone that spans the sample. This picture of the reconstruction is consistent with a model of the precursor phase due to Rinderer's group.(Danilov, Kubatkin, Landau, Parshin and Rinderer 1995; Danilov, Kubatkin, Landau and Rinderer 1995) They propose that the low temperature low-coverage phase is composed of an amorphous agglomeration of singlet dimers of the metal atoms. These molecules of metal atoms are metastable and have an activation barrier that prevents them from joining with other dimers or atoms to form larger clusters. Though a little vague on the details this model offers a plausible explanation of the insulating state.

There is also support for a highly disordered low temperature phase from *in situ* x-ray diffraction data taken on films while still in a specialy designed growth chamber.(Wissmann and Finzel 2007) At 87° K films show no (111) reflections but at 188° K these are present. Analysis using Sherrer's formula, which relates the width of the x-ray diffraction peaks to the average grain size, shows no discernible crystallites at 77° K but the onset of crystallites whose dimensions are smaller than the film thickness appears after annealing to 150° K.

As far as we know, low temperature, high resolution STM studies have not been made to characterize the microstructure of the precursor phase in silver. Published work for films of Gold on highly oriented pyrolitic graphite (HOPG) show a homogeneous coverage up to a 1.6 nm thickness, but the formation of islands for a thickness of 1.8 nm or greater.(Ekinci and Jr. 1998; Ekinci and Jr. 1998) The authors of this work have results showing that the behavior of Silver is qualitatively similar.(J. M. Valles 2000) Below a certain thickness a homogeneous phase forms, crossing over to crystallite formation as more material is added. This is consistent with the picture of a homogeneous precursor phase in our samples.

Our previous work has shown that if there is significant spatial inhomogeneity in thin films of silver then it can be observed by an enhancement of the temperature dependence of resistivity. This enhancement is a consequence of the assumption of a uniform sample at a given thickness that goes into the calculation of resistivity from the measured resistance data. The inhomogeneity needs to be accounted for by a scaling factor that takes into account the fractal geometric structure of the film. Conversely one can infer the inhomogeneity from the renormalization of the resistivity. While there are a number of ways of enhancing the residual resistivity, for example impurities or grain structure, the increased slope of the temperature dependence of resistivity is a conclusive signature of spatial inhomogeneity.

Measuring the resistivity of the films ex situ, after growth, can give significant insight into the films homogeneity. Figure 4 shows resistivity as a function of temperature for three thin films of silver. The lowest resistivity film is the same film whose resistance during growth is shown in figure 1. It has been grown to a thickness where it is homogeneous and while it has enhanced residual resistivity it has a temperature dependence that is indistinguishable from that of bulk samples of silver. The middle curve is a percolating silver film 320 angstroms thick. It clearly shows the geometric enhancement. Finally, the top curve is a quench condensed film 28 angstroms thick. It has the same temperature dependence as the homogeneous film, though its residual resistivity is enhanced. Two-step growth on glass produces homogeneous films though the enhancement of the residual resistivity at low temperatures indicates that there these films are clearly very disordered.

Another indication of this structural homogeneity can be seen in the low temperature magnetoresistance of these films. While the inhomogeneous films show a peculiar negative magnetoresistance, that is a consequence of the tortuous current paths in the percolating geometry, with a modification of the magnetoresistance due to weak localization at low fields, the two step grown films show clear two dimensional weak localization without any modifications.(Arnason and Hebard 1998) Again, the transport measurements are consistent with homogeneous two-dimensional films.

The "two-step" films on glass are metastable. If you raise their temperature to $T_a$ they reconstruct into discontinuous electrically isolated grains, indistinguishable from films of the same thickness grown at room temperature. This is not simply a question of speeding up an annealing that is occurring at room temperature. At room temperature, in vacuum, the films are stable for weeks without significant change in their structural or electrical properties. This second characteristic temperature seems to correlate with the energy barrier that is stabilizing the thin films. In the electronic growth model this energy is a combination of the change in the total energy of the silver due to both quantum confinement and the leakage of electron density into the substrate. Glass should present a significant barrier to electron leakage. This probably explains why the films reconstruct at a significantly lower temperature that has been seen for GaAs (110) where de-wetting occurs at 670° K(Yu, Jiang, Ebert, Wang, White, Niu, Zhang and Shih 2002), though it is surprising that these films are as stable as they are.

**Conclusion;**

We have presented data for ultra-thin films of Silver grown on glass by the "two-step" process. By measuring the resistance of the films, in situ, during the growth process, we have been able to observe that the precursor phase that anneals into the ultra thin film has a peculiar, insulating nature. Furthermore, the transition from precursor to metallic Silver occurs at a characteristic temperature. The insulating precursor phase is consistent with the singlet dimer model proposed by Rinderer. This suggests that the characteristic reconstruction temperature might be identifiable with the energy barrier that must be overcome to form larger clusters. Measuring the temperature dependence of resistivity and low temperature magnetoresistance *ex situ*, shows that these films are two-dimensional and extremely homogeneous. Finally, the films revert to their equilibrium microstructure at a temperature of 350° K. This temperature is indicative of the energy scale of the mechanism that is stabilizing the ultra-thin films. This stabilization might be due to the electronic growth model though it would be surprising that this model would predict such a large kinetic barrier for films of silver grown on glass.

**References;**

**Figure Captions**

Figure 1) Sheet resistance as a function of thickness during growth for a quench condensed film and a film grown at room temperature. The quench-condensed film, in the lower frame, shows a monotonic decrease in resistance as a function of thickness. At approximately 330 Å the slope changes when the deposition rate was doubled. The room temperature film, in the upper frame, shows exponential behavior during the nucleation due to hopping of electrons between isolated clusters. This crosses over to power law dependence at about 180 Å as the clusters percolate. Eventually, the resistance varies as 1/thickness when the film becomes homogeneous in the plane of the substrate, at about 500Å.

Figure 2) Sheet resistance as a function of temperature as a nominal coverage of 43 Å of quench condensed Ag anneals to room temperature, measured *in situ*. At a characteristic temperature, ~150K, the film dramatically reconstructs and its resistance drops by 9 orders of magnitude. Inset shows a similar film where the reconstruction has been suppressed by rapid cooling. For R > 20 kΩ the resistance is activated. For R < 20 kΩ the temperature dependence is metallic.

Figure 3) Sheet resistance as a function of temperature as a nominal coverage of 28 Å of quench condensed Ag anneals to room temperature, measured *in situ*. As in figure 2 the sample reconstructs at a characteristic temperature but here we quench the reconstruction by rapidly cooling the sample. For sheet resistances greater the $10^4$ ohms the resistance as a function of temperature is activated. Below $10^4$ ohms it is metallic. The film continues to anneal up to 280° K above which the resistance is again reversible in temperature.

Figure 4) Resistivity as a function of temperature for three silver films grown on glass substrates, measured *ex situ*, after growth. The Bottom most set of data, triangular markers, is for a film grown at room temperature to a thickness where it has become homogeneous, having the temperature dependence of bulk silver and enhanced residual resistivity due to grain boundary scattering. The middle set of data, square markers, is for a film grown at room temperature, but with the growth terminated during the percolation of the film. This film shows enhancement of both the residual resistance and the slope of the temperature dependence of resistance as a consequence of the inhomogeneous microstructure. This purely geometric factor is a multiplicative constant that collapses the data, onto the lower curve. The topmost set of data, circular markers, shows resistivity as a function of temperature for 50 Å of silver grown on glass by the two-step process. Its temperature dependence is that of bulk silver, showing that the film is homogeneous. But, there is significantly enhanced residual resistivity due to disorder.

Figure 1

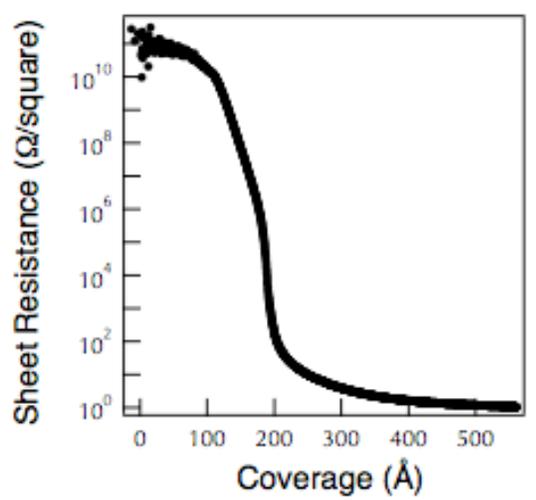
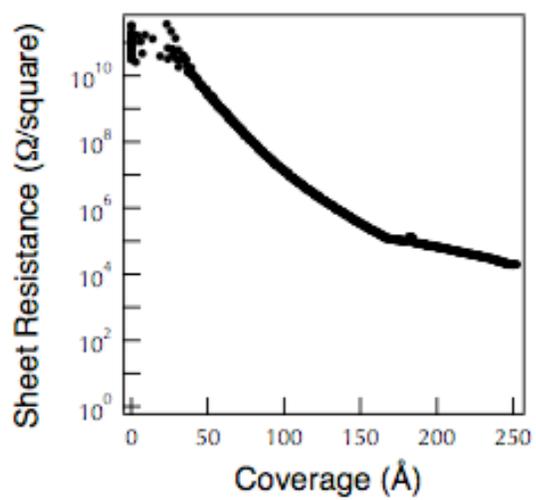

Figure 2.

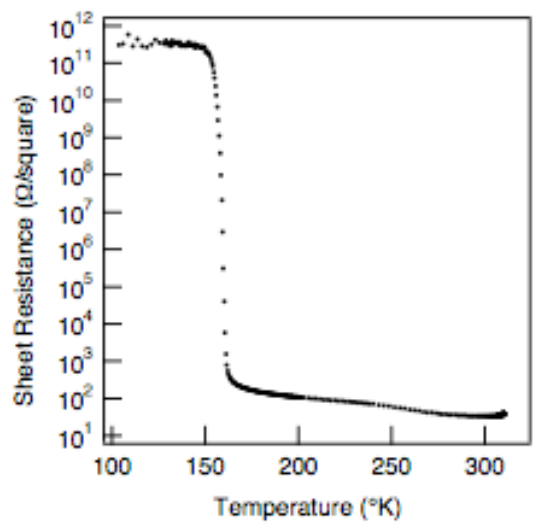

Figure 3.

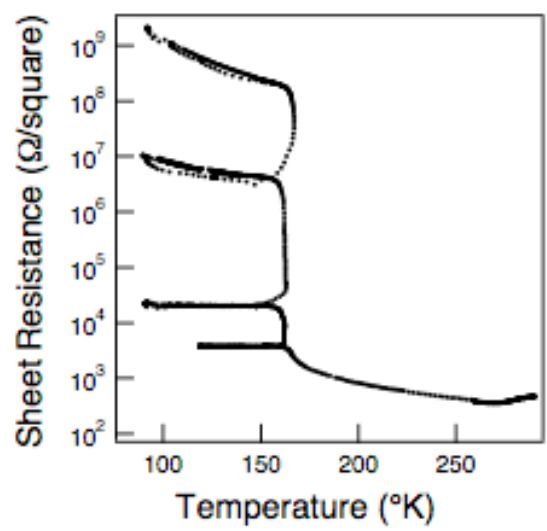

Figure 4.

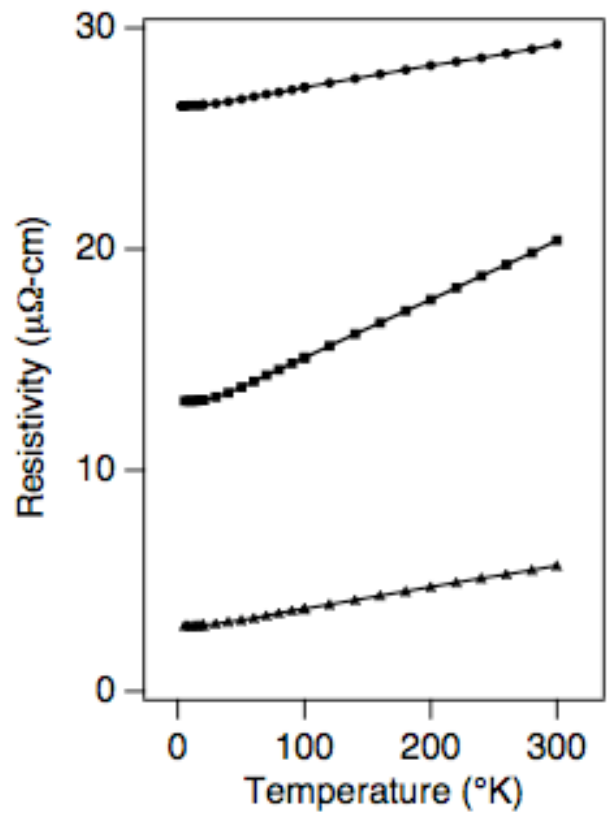